\documentclass[lettersize,journal]{IEEEtran}
\usepackage{amsmath,amsfonts}
\usepackage{algorithmic}
\usepackage{algorithm}
\usepackage{array}
\usepackage[caption=false,font=normalsize,labelfont=sf,textfont=sf]{subfig}
\usepackage{textcomp}
\usepackage{stfloats}
\usepackage{url}
\usepackage{verbatim}
\usepackage{graphicx}
\usepackage{cite}

\usepackage{hyperref}
\usepackage{multirow}
\usepackage{bbding}
\usepackage{booktabs}
\usepackage{amssymb}

\begin{document}

\title{Retrosynthesis Prediction via \\ Search in (Hyper) Graph}

\author{Zixun Lan$^{1}$ \qquad Binjie Hong$^{2}$ \qquad Jiajun Zhu$^{3}$ \qquad Zuo Zeng$^{4}$  \qquad  Zhenfu Liu$^{4}$ \qquad  Limin Yu$^{3}$ \qquad
Fei Ma$^{1*}$\\
$^1$\ Department of Applied Mathematics, Xi’an Jiaotong-Liverpool University\\
$^2$\ Department of Computer Science, The University of Manchester\\
$^3$\ Electrical and Electronic Engineering, Xi’an Jiaotong-Liverpool University\\
$^4$\ Hours Technology Co., Ltd., Suzhou, China\\
\{zixun.lan19, jiajun.zhu22\}@student.xjtlu.edu.cn, binjie.hong@postgrad.manchester.ac.uk, \\
\{zuo.zeng, liuzhenfu\}@woshikeji.net, \{limin.yu, fei.ma\}@xjtlu.edu.cn
\thanks{$^*$corresponding author}
}

\markboth{Journal of \LaTeX\ Class Files,~Vol.~14, No.~8, August~2021}%
{Shell \MakeLowercase{\textit{et al.}}: A Sample Article Using IEEEtran.cls for IEEE Journals}


\maketitle

\begin{abstract}
Predicting reactants from a specified core product stands as a fundamental challenge within organic synthesis, termed retrosynthesis prediction. Recently, semi-template-based methods and graph-edits-based methods have achieved good performance in terms of both interpretability and accuracy. However, due to their mechanisms these methods cannot predict complex reactions, e.g., reactions with multiple reaction center or attaching the same leaving group to more than one atom. In this study we propose a semi-template-based method, the \textbf{Retro}synthesis via \textbf{S}earch \textbf{i}n (Hyper) \textbf{G}raph (RetroSiG) framework to alleviate these limitations. In the proposed method, we turn the reaction center identification and the leaving group completion tasks as tasks of searching in the product molecular graph and leaving group hypergraph respectively. As a semi-template-based method RetroSiG has several advantages. First, RetroSiG is able to handle the complex reactions mentioned above by its novel search mechanism. Second, RetroSiG naturally exploits the hypergraph to model the implicit dependencies between leaving groups. Third, RetroSiG makes full use of the prior, i.e., one-hop constraint. It reduces the search space and enhances overall performance. Comprehensive experiments demonstrated that RetroSiG achieved competitive results. Furthermore, we conducted experiments to show the capability of RetroSiG in predicting complex reactions. Ablation experiments verified the efficacy of specific elements, such as the one-hop constraint and the leaving group hypergraph.
\end{abstract}

\begin{IEEEkeywords}
Retrosynthesis, Reinforcement Learning, Semi-Template-Based Method.
\end{IEEEkeywords}

\section{Introduction}
\IEEEPARstart{T}{he} retrosynthesis prediction aims to discern the appropriate reactants necessary for synthesizing a designated product molecule. It was initially formalized by \cite{corey1991logic} and has since evolved into a foundational issue within the realm of organic chemistry. The challenge lies in the vastness of the search space, encompassing a multitude of potential transformations, as noted by \cite{gothard2012rewiring}. Thus, there has been a quest for a range of computer algorithms aimed at supporting seasoned chemists in their endeavors. Among them, machine learning based methods have played a vital role and achieved significant progress in this area recently \cite{coley2017computer}.

The template-based and template-free methods have their own set of advantages and disadvantages. The template-based methodologies \cite{coley2017computer, dai2019retrosynthesis, segler2017neural, chen2021deep, yan2022retrocomposer} tackle retrosynthesis prediction as a template retrieval problem. Following template retrieval, these approaches utilize cheminformatics tools like RDKit \cite{landrum2006rdkit} to construct full reactions. Despite their high interpretability and assurance of molecule validity, they cannot predict reactions outside the template library. In contrast, the template-free approaches leverage deep generative models to produce reactants based on a given product directly. Since molecules can be represented using both graphs and SMILES sequences, existing techniques reframe retrosynthesis as either a sequence-to-sequence problem \cite{lin2020automatic, zheng2019predicting, tetko2020state, seo2021gta, wan2022retroformer} or a graph-to-sequence problem \cite{tu2022permutation}. Although these generative methods facilitate chemical reasoning within a more expansive reaction space, they lack interpretability.

\begin{figure*}
	\centering
	\includegraphics[width=1\linewidth]{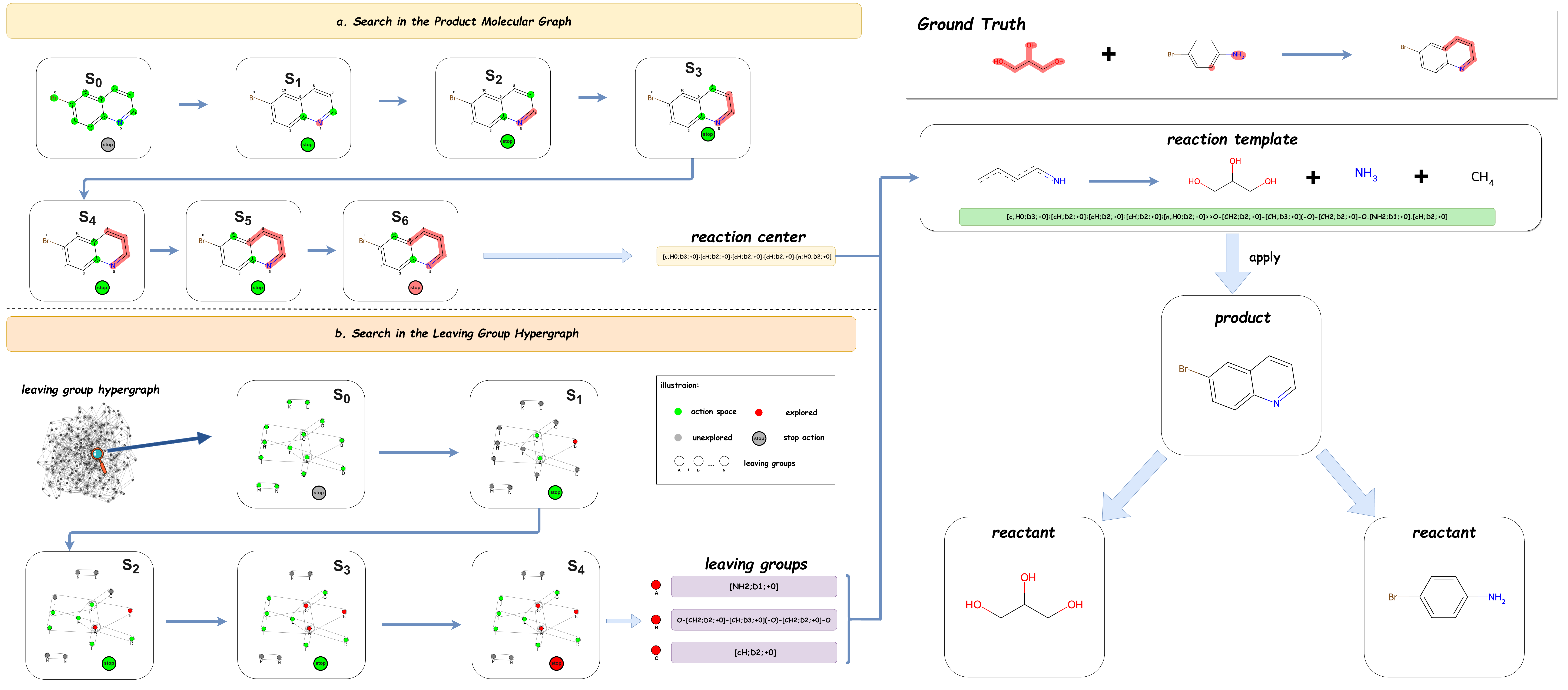}
	\caption{Architecture overview. RretroSiG first identifies reaction center via \textbf{(a)} search in the product molecular graph and then completes leaving groups through \textbf{(b)} search in the leaving group hypergraph. Finally, RetroSiG converts predicted subgraph into reaction center SMARTS and leaving groups SMARTS, subsequently merging them to derive one retrosynthesis template. RetroSiG obtains the reactants by applying the merged template to the given product. At state $t$, the red highlighted part represents the explored nodes, and the green nodes denote the action space after applying the one-hop constraint.}
	\label{fig1}
\end{figure*}

The semi-template approaches effectively leverage the advantages inherent in generative models along with established chemical knowledge to enhance their methodology. Typical frameworks \cite{yan2020retroxpert, shi2020graph, somnath2021learning, wang2021retroprime, chen2023g2retro} within this category adhere to a common strategy: they first pinpoint the reaction center and convert the product into synthons. Then, a separate model completes the transformation of these synthons into reactants. These methods offer competitive accuracy and maintain interpretability due to their stage-wise nature. Nonetheless, they continue to face challenges in predicting reactions with multiple reaction center and fail in the scenario where the same leaving group is attached to more than one atom in a molecular graph. This paper refers to the reaction center comprised of multiple bonds or atoms as multiple raction center. Fig. \ref{fig1} shows a reaction with multiple reaction center and attaching the same leaving group to more than one atom.

Recently, some methodologies \cite{sacha2021molecule, zhong2023retrosynthesis} have framed retrosynthesis prediction as the task of editing the product molecular graph to the reactant molecular graph. In this paper, we refer to this type of method as the graph edit sequence conditional generative model. The edit operations encompass actions such as Delete-bond, Change-bond, Change-atom, Attach-leaving-group, etc. These techniques extract edit sequences from ground-truth reaction data and then model the conditional probability distribution over these sequences given one product molecular graph. Despite its mitigation of some template-based method limitations, the graph edit sequence conditional generative model remains challenged. It cannot handle attaching the same leaving group to more than one atom in a molecular graph and also underutilizes the prior of chemical rules.

Motivated by the aforementioned constraints, we introduce the RetroSiG (\textbf{Retro}synthesis via \textbf{S}earch \textbf{i}n (Hyper) \textbf{G}raph) framework (Fig. \ref{fig1})—a semi-template-based approach—aimed at facilitating single-step retrosynthesis prediction. There are two phases in our RetroSiG, i.e., (a) search in the product molecular graph and (b) search in the leaving group hypergraph. \textbf{(a) (Fig. \ref{fig1}a)}: We regard the reaction center identification as the search in the product molecular graph. A crucial observation dictates that the single or multiple reaction center should constitute a node-induced subgraph within the molecular product graph \cite{coley2019rdchiral}. Therefore, RetroSiG utilizes a reinforcement learning agent with the graph encoder to explore the appropriate node-induced subgraph in the specific product molecular graph, identifying it as the predicted reaction center. At each stage, it contemplates the selection of one node within the molecular product graph as an action, incorporating it into the explored node-induced subgraph. As the reaction center graph maintains connectivity, we implement a one-hop constraint to streamline the search space and optimize the overall performance. \textbf{(b) (Fig. \ref{fig1}b)}: We regard the leaving group completion as the search in the leaving group hypergraph. Since the unique leaving group patterns derived from the training dataset can cover 99.7\% of the test set \cite{somnath2021learning}, it is feasible to predict leaving group patterns from the training leaving group corpus. Implicit dependencies exist among various leaving groups. Naturally, we can model these dependencies by creating a leaving group hypergraph. Each node denotes one unique leaving group pattern, while a hyperedge connects the corresponding leaving groups that co-occur within a template (reaction). Therefore, RetroSiG also utilizes a reinforcement learning agent with the hypergraph encoder to explore the appropriate node-induced subgraph in the leaving group hypergraph, selecting it as the resulting leaving groups. At each step, it contemplates the selection of one node within the leaving group hypergraph as an action, incorporating it into the explored node-induced subgraph. In the evaluation set, we observe that 98.9\% of the leaving groups that co-occur within a template are contained within the same hyperedges. Thus, we make full use of the prior and impose a one-hop constraint to reduce the search space. \textbf{Finally}, we convert two predicted node-induced subgraphs into reaction center SMARTS and leaving groups SMARTS, subsequently merging them to derive one retrosynthesis template. We obtain the reactants by applying the retrosynthesis template to the given product \cite{coley2019rdchiral}. Comprehensive experiments demonstrate that RetroSiG achieves a competitive result. Furthermore, we conduct experiments to show the capability of RetroSiG in predicting complex reactions. Ablation experiments confirm the efficacy of specific elements, such as the one-hop constraint and the leaving group hypergraph. In summary, we outline our primary contributions as follows:
\begin{itemize}
\item We propose, a semi-template-based method, the RetroSiG framework to conduct single-step retrosynthesis prediction. It has the capability to predicting reactions with multiple reaction center and attaching the same leaving group to more than one atom.

\item The key novelty of RetroSiG is viewing the reaction center identification and leaving group completion as the search in the product molecular graph and the leaving group hypergraph respectively. The hypergraph effectively represents the implicit dependencies among leaving groups, and the search in the graph makes full use of the prior, i.e., one-hop constraint.

\item Comprehensive experiments demonstrate that RetroSiG achieves a competitive result. Furthermore, we conduct experiments to show the capability of RetroSiG in predicting complex reactions. Ablation experiments confirm the efficacy of specific elements.
\end{itemize}

\section{Related Work}

We summarize the comparison of existing single-step retrosynthesis in Table \ref{related work}. Compared with TB, RetroSiG does not need to perform \textbf{Subgraph Matching} (NP-Hard) since we can determine the exact location of the reaction center by the explored node-induced subgraph. Also, the specific position leads to the unique solution after applying the template, and thus RetroSiG does not need the additional model to \textbf{ranking Reactants}. RetroSiG can obtain \textbf{Explicit Template} and have the capability of predicting complex reactions with \textbf{multiple reaction center (RC)} or \textbf{attaching the same leaving group (LG) to more than one atom}.

\begin{table*}
\label{related work}
\caption{Comparison of different baselines in seven dimensions.}
\begin{center}

\resizebox{\textwidth}{!}{%
    \begin{tabular}{l|ccccccc}
\toprule
\multirow{2}{*}{} & \multirow{2}{*}{\shortstack{Interpretability}} & \multirow{2}{*}{\shortstack{Subgraph\\Matching}} & \multirow{2}{*}{\shortstack{Ranking\\Reactants}} & \multirow{2}{*}{\shortstack{Explicit\\Template}} & \multirow{2}{*}{\shortstack{Multi-RC}} & \multirow{2}{*}{\shortstack{Attaching the Same LG \\ to More Than One Atom}}  & \multirow{2}{*}{\shortstack{Extrapolation\\Ability}}                                             \\
                  &                                   &                                    &                                    &                                    &                                                       &  &                                                                                    \\ \midrule
TB                &                  \CheckmarkBold                &                  \CheckmarkBold                  &   \CheckmarkBold                                  &      \CheckmarkBold                               &      \CheckmarkBold                                                  &    \CheckmarkBold   & \XSolidBrush                                                                                \\
TF                &  \XSolidBrush                                 &  \XSolidBrush                                  & \XSolidBrush                                   &   \XSolidBrush                                 &     \CheckmarkBold                                                &  \CheckmarkBold  &  \CheckmarkBold                                                                                     \\
STB               &      \CheckmarkBold                              &     \XSolidBrush                               &    \XSolidBrush                                &                         \CheckmarkBold          &    \XSolidBrush                                                   &            \XSolidBrush    &  \CheckmarkBold                                                                        \\
Graph2Edits       &    \CheckmarkBold                               &                  \XSolidBrush                  &                               \XSolidBrush     &                           \CheckmarkBold        &  \CheckmarkBold                                                      &  \XSolidBrush      &  \CheckmarkBold                                                                                \\
RetroSiG          &        \CheckmarkBold                            &   \XSolidBrush                                 &   \XSolidBrush                                 &         \CheckmarkBold                            &  \CheckmarkBold                                                      &                             \CheckmarkBold       &  \CheckmarkBold                                                     \\ 
\bottomrule
\end{tabular}
  }
  
\end{center}
\end{table*}

\subsection{Graph Edit Sequence Conditional Generative Methods} MEGAN \cite{sacha2021molecule} and Graph2Edits \cite{zhong2023retrosynthesis} express a chemical reaction as a sequence of graph edits $\boldsymbol{E}=\left(e_0, \cdots, e_t, \cdots, e_T\right)$ that transform the product molecular graph into the reactant molecular graph. Here, the edits $e$ encompass various actions such as Delete-bond, Change-bond, Change-atom, Attach-leaving-group, and so forth. The probability of a sequence of graph edits $\boldsymbol{E}=\left(e_0, \cdots, e_t, \cdots, e_T\right)$ can be factorized to the probabilities conditional on the product molecule $\boldsymbol{p}$ and $\boldsymbol{E}_{<t}=\left(e_0, \cdots, e_{t-1}\right)$, i.e., 
\begin{equation}
p(\boldsymbol{E} \mid \boldsymbol{p})=\prod_{t=1}^T p\left(e_t \mid \boldsymbol{p}, \boldsymbol{E}_{<t}\right), T \geq 1
\end{equation}
When $T = 0$, $\boldsymbol{E}=\left(e_0\right)$ and $p(e_0 \mid \boldsymbol{p}) = 1$ because $e_0$ is a fixed start token. These techniques extract edit sequences from ground-truth reaction data and then model the conditional probability distribution over these edit sequences given one product molecular graph. MEGAN \cite{sacha2021molecule} was the initial attempt to represent reactions as edits sequences for retrosynthesis prediction. However, this approach faced difficulties in reactant generation due to the atomic-level addition operations. It particularly struggled with reactions involving the attachment of large leaving groups. Graph2Edits \cite{zhong2023retrosynthesis} presents retrosynthesis as a process of reasoning involving reactions between products, intermediates, and reactants, achieved through a sequence of graph modifications. Importantly, in contrast to the MEGAN model, Graph2Edits substitutes the "add-atom" actions with the attachment of substructures, reducing the number of generation steps and thereby enhancing reactant generation efficiency. These methods acquire knowledge of reaction transformation rules to a certain degree, thus improving their interpretability and generalization capabilities in complex reactions. However, It cannot predict reactions with attaching the same leaving group to more than one atom in a molecular graph and also underutilizes the prior of chemical rules.

\subsection{Template-based Methods (TB)} The template serves as a reaction representation that's both more efficient and interpretable \cite{coley2019rdchiral}. After establishing a library of reaction templates, the methods involve comparing a target molecule against these templates. Upon finding a match, the corresponding template is utilized to transform product molecules into reactant molecules. Numerous research efforts have introduced various strategies for prioritizing templates. In prior studies, \cite{coley2017computer} utilized molecular fingerprint similarity between the intended product and compounds in the dataset to prioritize potential templates. Meanwhile, \cite{segler2017neural} employed Neuralsym, a hybrid neural-symbolic model, to glean insights for a multi-class classification task focused on template selection. Additionally, \cite{dai2019retrosynthesis} interpreted chemical insights within reaction templates as logical rules, utilizing graph embeddings to derive the conditional joint probability of these rules and reactants. Furthermore, \cite{chen2021deep} evaluated pertinent local templates within forecasted reaction centers of a target molecule. It also accounted for the non-local impacts of chemical reactions by integrating global reactivity attention. \cite{yan2022retrocomposer, zhu2023single} composes templates by selecting and annotating molecule subgraphs from training templates. Despite their interpretability, template-based methods have restricted applicability due to their incapability to forecast reactions beyond the template library. Moreover, their suitability diminishes when dealing with extensive template sets due to the challenges inherent in subgraph matching.

\subsection{Template-free Methods (TF)} Template-free Methods utilize deep generative models to generate reactant molecules directly. Prior research \cite{lin2020automatic, zheng2019predicting, tetko2020state, seo2021gta, kim2021valid, wan2022retroformer} has employed the Transformer architecture \cite{vaswani2017attention}, reconceptualizing the issue as a sequence-to-sequence translation task, transforming from product molecules to corresponding reactants. \cite{tu2022permutation} replaces the initial sequence encoder with a graph encoder to guarantee the permutation invariance of SMILES representations. Although these generative methods facilitate chemical reasoning within a more expansive reaction space, they lack interpretability.

\subsection{Semi-template-based Methods (STB)} Semi-template-based approaches amalgamate the advantages of generative models with additional chemical knowledge. Considering that chemical reactions generally entail altering a limited segment of the molecular structure, most existing research approaches have divided retrosynthesis into a two-stage procedure: first, identifying the reaction center utilizing a graph neural network to generate synthons via molecular editing \cite{lan2023rcsearcher}, and subsequently converting these synthons into reactants employing methods like a graph generative model \cite{shi2020graph}, a Transformer \cite{shi2020graph, wang2021retroprime}, or a subgraph selection model \cite{somnath2021learning}. These methods offer competitive accuracy and maintain interpretability due to their stage-wise nature. Nonetheless, they continue to face challenges in predicting reactions with multiple reaction center. They cannot effectively address scenarios where the same leaving group is attached to more than one atom in a molecular graph \cite{lan2021sub, lan2023aednet, lan2022more, lan2023sub}. 

\section{Methodology}
\subsection{Preliminaries}
\label{preliminaries}
\subsubsection{Edge Graph Attention Network Layer}
Given $\mathbf{H}^{(t)}= [ \boldsymbol{h}_{1}^{(t)}; \boldsymbol{h}_{2}^{(t)}; \cdots ; \boldsymbol{h}_{n}^{(t)}] \in R^{n \times d}$ and $\mathbf{f}_{i j}^{(t)}\in R^{1 \times d}$ are the node embedding matrix and the embedding of edge $(i, j)$ at the $t$-th layer repectively, where $\boldsymbol{h}_{i}^{(t)}\in R^{1 \times d}$ is the node-level embedding for node $i$ of the graph and is also the $i$-th row of $\mathbf{H}^{(t)}$, $d$ is the dimension of node-level embedding and $n$ is the number of nodes. $\boldsymbol{h}_{i}^{(0)}$ and $\boldsymbol{f}_{ij}^{(0)}$ are the initial features of node and edge in molecular product graph $\mathcal{G}_{p}$. EGAT injects  the graph structure into the attention mechanism by performing masked attention, namely it only computes $\alpha_{i j}$ for nodes $j \in \mathcal{N}_{i}$, where $\mathcal{N}_{i}$ is the first-order neighbors of node $i$ in the graph:
\begin{equation}
	\label{egat}
	\begin{aligned}
        \boldsymbol{f}_{i j}^{(t+1)}&=\text { LeakyReLU }\left(\left[\boldsymbol{h}_i^{(t)}\mathbf{W}\left\|\boldsymbol{f}_{i j}^{(t)}\right\| \boldsymbol{h}_j^{(t)}\mathbf{W}\right]A\right), \\
		e_{ij} &= \mathbf{a} \cdot {\boldsymbol{f}_{i j}^{(t+1)}}^{T}, \alpha_{i j} =\frac{\exp \left(e_{ij}\right)}{\sum_{k \in \mathcal{N}_{i}} \exp \left(e_{ik}\right)},
	\end{aligned}
\end{equation}
where $e_{ij}\in R$ and $\alpha_{i j}\in R$ are a non-normalized attention coefficient and a normalized attention coefficient representing the weight of message aggregated from node $j$ to node $i$ respectively in the $t$-th layer of EGAT, and $\|$ is the concatenation operation. Besides,$\mathbf{W}\in R^{d \times d}$, $A\in R^{3d \times d}$ and $\mathbf{a}\in R^{1 \times d}$ are learnable parameters in the $t$-th layer.

EGAT \cite{kaminski2022rossmann} employs multi-head attention to stabilize the learning process of self-attention, similar to Transformer \cite{vaswani2017attention}. If there are $K$ heads, $K$ independent attention mechanisms execute the Eq. \ref{egat}, and then their features are concatenated:
\begin{equation}
	\label{att}
	\begin{aligned}
	    \boldsymbol{h}_{i}^{(t+1)} &= \operatorname{MLP} \left(\|_{k=1}^{K} \sigma\left(\sum_{j \in \mathcal{N}_{i}} \alpha_{i j}^{k}  \boldsymbol{h}_{j}^{(t)} \mathbf{W}^{k}\right)\right)
	\end{aligned}
\end{equation}
where $\|$ represents concatenation, $\alpha_{i j}^{k}$ are normalized attention coefficients computed by the $k$-th learnable $\mathbf{W}^{k}\in R^{d \times d}$, $A^{k}\in R^{3d \times d}$ and $\mathbf{a}^{k}\in R^{1 \times d}$ following Eq. \ref{egat}. Besides, $\operatorname{MLP}$ denotes multi-perceptron.

\subsubsection{Hyper Graph Neural Network Layer}
Specifically, for the $l$-th layer in HGNN \cite{feng2019hypergraph}, it takes hypergraph $\mathcal{G}_{hg}$’s incidence matrix $\mathrm{H} \in R^{n \times e}$ and hidden representation matrix $\mathrm{X}^l \in R^{n \times d^{l}}$ as input, where $n$ and $e$ are the number of nodes and hyper-edges respectively. Then, the node representations in next layer will be computed as follows:
\begin{equation}
\begin{aligned}
\mathbf{X}^{l+1}=\sigma\left(\mathbf{D}_v^{-\frac{1}{2}} \mathbf{H W D}_e^{-1} \mathbf{H}^{\top} \mathbf{D}_v^{-\frac{1}{2}} \mathbf{X}^l \boldsymbol{\Theta}^l\right)
\end{aligned}
\end{equation}
where $\sigma(\cdot)$ is the nonlinear activation function. $\mathbf{D}_v \in R^{n \times n}, \mathbf{D}_e \in R^{e \times e}, \mathbf{W} \in R^{e \times e}$ are the diagonal node degree, edge degree and edge weight matrices, respectively. $\mathbf{\Theta}^l \in R^{d^{l} \times d^{l+1}}$ is a trainable parameter matrix.

\subsection{Search in the Product Molecular Graph}
In this study, RDChiral \cite{coley2019rdchiral} is employed to identify the super general reaction center. Consequently, it ensures that whether it's a single or multiple reaction center, it must be a node-induced subgraph within the molecular product graph. A product molecular graph $\mathcal{G}_{p}=(\mathcal{V}_{p}, \mathcal{E}_{p})$ is represented as a set of $|\mathcal{V}_{p}|$ nodes (atoms) and a set of $|\mathcal{E}_{p}|$ edges (bonds). The reaction center graph $\mathcal{G}_{rc}=(\mathcal{V}_{rc}, \mathcal{E}_{rc})$ is a node-induced subgraph of the product molecular graph $\mathcal{G}_{p}$ such that $\mathcal{V}_{rc} \subseteq \mathcal{V}_{p}$ and $\mathcal{E}_{rc}=\left\{(u, v) \mid u, v \in \mathcal{V}_{rc},(u, v) \in \mathcal{E}_{p}\right\}$. Given the product molecular graph $\mathcal{G}_{p}$, reaction center identification aims to detect the corresponding reaction center graph $\mathcal{G}_{rc}$, i.e. the node set $\mathcal{V}_{rc}$ of $\mathcal{G}_{rc}$. Thus, we regard the reaction center identification as the search in the product molecular graph. At each stage, RetroSiG deliberates on selecting a node within the molecular product graph as an action, incorporating it into the explored node-induced subgraph. Ultimately, we transform the identified subgraph into the SMARTS of reaction center.

\textbf{State Space:} RetroSiG has a discrete state space $\mathcal{S}$. Each state $s_t \in S$ is represented as $s_t = \{\mathcal{G}_{p}, \hat{\mathcal{V}}_{rc}^{t}\}$, where $\hat{\mathcal{V}}_{rc}^{t}$ denotes the explored node set at step $t$. Notably, $\hat{\mathcal{V}}_{rc}^{t=0} = \emptyset$.

\textbf{Action Space:} RetroSiG has two types of actions in its action space $\mathcal{A}$: (1) selecting one node $v$ in the product molecular graph $\mathcal{G}_{p}$, and (2) stop action, (i.e., doing nothing), denoted as $\text{STOP}$. Since the reaction center graph $\mathcal{G}_{rc}$ is connected, we impose a one-hop constraint on the first type of action. This constraint serves to reduce the search space and enhance overall performance. In other words, the action $a_t$ is to select one node $v$ from the first-order neighbour of the current explored subgraph $\hat{\mathcal{G}}_{rc}^{t}$ or a $\text{STOP}$ action implying stop of the exploration process. At step $t$, action space is represented as $\mathcal{A}^{t>0} = \{v | v \in \operatorname{ONE-HOP}(\hat{\mathcal{G}}_{rc}^{t})\} \cup \{\text{STOP} \}$. Notably, $\mathcal{A}^{0} = \{v | v \in \mathcal{V}_{p} \}$.


\begin{figure*}
	\centering
	\includegraphics[width=1\linewidth]{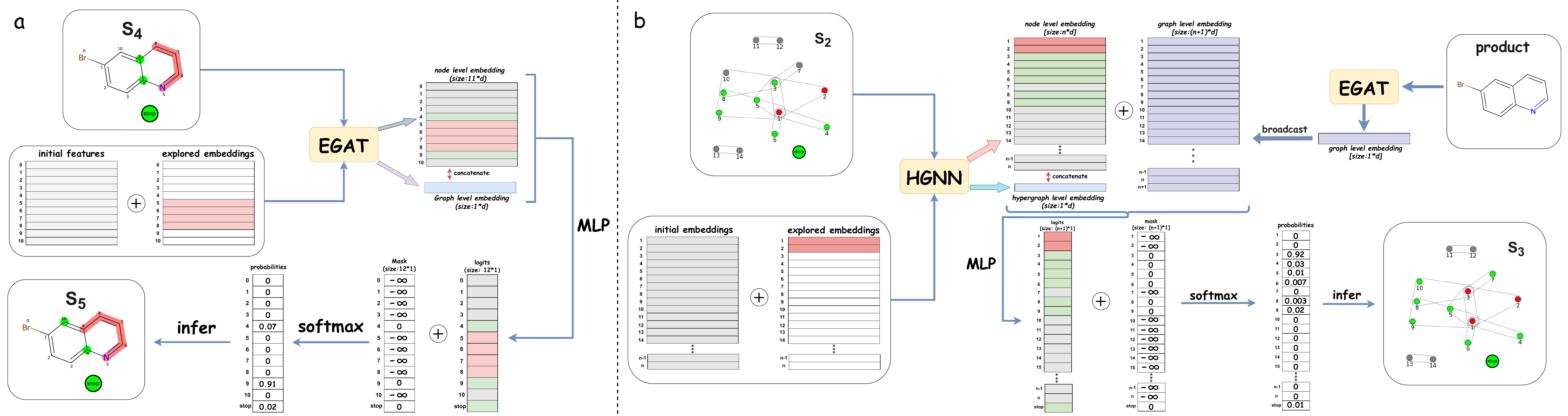}
	\caption{\textbf{a.} Policy network architecture in the search in the product molecular graph. \textbf{b.} Policy network architecture in the search in the leaving group hypergraph.}
	\label{policy}
\end{figure*}

\textbf{Reward Function:} At intermediate step $t$, reward function $\mathcal{R}$ gives $s_t$ a reward 0. At the terminal step $T$, if the predicted $\hat{\mathcal{V}}_{rc}^{T}$ is identical to the ground-truth $\mathcal{V}_{rc}$, $\mathcal{R}$ gives $s_t$ a reward 1. Otherwise, it gives $s_t$ a reward of 0.

\textbf{Policy Net (Fig. \ref{policy}a):} 
We use the EGAT \cite{kaminski2022rossmann} to obtain node-level embedding. Each atom $u$ has an initial feature vector $\mathbf{x}_u$. Each bond $(u, v)$ has a feature vector $\mathbf{x}_{u v}$. The specifics regarding the initial features can be located in Table \ref{atom} and Table \ref{bond}. For ease of reference, we symbolize the encoding process as $\mathrm{EGAT}(\cdot)$ and elaborate on the architectural specifics in Section \ref{preliminaries}. The EGAT computes atom representations $\left\{\mathbf{c}_u^{t} \mid u \in \mathcal{V}_p\right\}$ via
\begin{equation}
\resizebox{1\linewidth}{!}{
$\left\{\mathbf{c}_u^{t}\right\}=\operatorname{EGAT}\left(\mathcal{G}_{p},\left\{\mathbf{x}_u + \mathbb{I}[u \in \hat{\mathcal{V}}_{rc}^{t}] \mathbf{x}_{explored}\right\},\left\{\mathbf{x}_{u v}\right\}_{v \in \mathcal{N}(u)}\right).$}
\end{equation}
In order to incorporate information about the explored node set $\hat{\mathcal{V}}_{rc}^{t}$ at step $t$, we add a learnable embedding $\mathbf{x}_{explored}$ to the initial feature of the explored node. $\mathbb{I}[u \in \hat{\mathcal{V}}_{rc}^{t}] \mapsto\{0,1\}$ be the predicate that indicates whether atom $u$ is the element of $\hat{\mathcal{V}}_{rc}^{t}$. Here, $\mathcal{N}(u)$ refers to the neighboring atoms of atom $u$. The graph representation $\mathbf{c}_{\mathcal{G}}^t$ is formed by aggregating atom representations, i.e. $\mathbf{c}_{\mathcal{G}}^t=\sum_{u \in \mathcal{V}_p} \mathbf{c}_u^t$. Finally, we calculate the logits $s_u^t \in \mathbb{R}^1, s_s^t \in \mathbb{R}^1$ for selecting one atom $u$ and STOP action at each step $t$ through the fully connected layers:
\begin{equation}
\begin{aligned}
s_u^t & =\boldsymbol{W}_b\left(\sigma\left(\boldsymbol{W}_a \mathrm{c}_u^t+\boldsymbol{B}_a\right)\right)-\mathbb{I}\left[u \notin \mathcal{A}^t\right] \infty, \\
s_s^t & =\boldsymbol{W}_b\left(\sigma\left(\boldsymbol{W}_a \mathrm{c}_{\mathcal{G}}^t+\boldsymbol{B}_a\right)\right)-\mathbb{I}\left[\mathrm{STOP} \notin \mathcal{A}^t\right] \infty,
\end{aligned}
\end{equation}
where $\boldsymbol{W}_a$ and $\boldsymbol{W}_b$ are the weights and $\boldsymbol{B}_a$ is the bias. $\mathbb{I}[u \notin \mathcal{A}^{t}] \, (\mathbb{I}[\text{STOP} \notin \mathcal{A}^{t}]) \mapsto\{0,1\}$ be the predicate that indicates whether atom $u$ ($\text{STOP}$) is not in the action space $\mathcal{A}^{t}$ at step $t$. We add the $-\infty$ to the logit of the action not appearing in the action space $\mathcal{A}^{t}$ at step $t$. It ensures that the probability of an action not appearing in the action space equals 0 after $\mathrm{SOFTMAX}$ operation.

\begin{table}[h]
\centering
\caption{Atom Features used in EGAT. All features are one-hot encoding.}
\label{atom}
\begin{tabular}{lcr}
\hline
\multicolumn{1}{c}{Feature} & Description                                    & \multicolumn{1}{c}{Size} \\ \hline
Atom type                   & Type of an atom by atomic number.              & 100                      \\
Total degree                & Degree of an atom including Hs.                & 6                        \\
Explicit valence            & Explicit valence of an atom.                   & 6                        \\
Implicit valence            & Explicit valence of an atom.                   & 6                        \\
Hybridization               & sp, sp2, sp3, sp3d, or sp3d2.                  & 5                        \\
\# Hs                       & Number of bonded Hydrogen atom.                & 5                        \\
Formal charge               & Integer electronic charge assigned to atom.    & 5                        \\
Aromaticity                 & Whether atom is part of aromatic system. & 1                        \\
In ring                 & Whether an atom is in ring                     & 1                        \\ \hline
\end{tabular}
\end{table}

\begin{table}[h]
\centering
\caption{Bond features used in EGAT. All features are one-hot encoding.}
\label{bond}
\begin{tabular}{lcr}
\hline
\multicolumn{1}{c}{Feature} & Description                          & \multicolumn{1}{c}{Size} \\ \hline
Bond type                   & Single, double, triple, or aromatic. & 4                        \\
Conjugation                 & Whether the bond is conjugated.      & 1                        \\
In ring                     & Whether the bond is part of a ring.  & 1                        \\
Stereo                      & None, any, E/Z or cis/trans.         & 6                        \\
Direction                   & The direction of the bond.             & 3                        \\ \hline
\end{tabular}
\end{table}

\subsection{Search in the Leaving Group Hypergrah}
In this paper, we use RDChiral \cite{coley2019rdchiral} to extract the super general leaving groups from the training set and obtain unique leaving group patterns. Since the unique leaving group patterns derived from the training dataset can cover 99.7\% of the test set \cite{somnath2021learning}, it is feasible to predict leaving group patterns from the training leaving group corpus. Implicit dependencies exist among various leaving groups. Naturally, we build a hypergraph to model these dependencies. Each node denotes one unique leaving group pattern, while a hyperedge connects the corresponding leaving groups that co-occur within a template. A leaving group hypergraph $\mathcal{G}_{hg}=(\mathcal{V}_{hg}, \mathcal{E}_{hg})$ is represented as a set of $|\mathcal{V}_{hg}|$ nodes (unique leaving group) and a set of $|\mathcal{E}_{hg}|$ hyperedges (co-occurrence). Given the product molecular graph $\mathcal{G}_{p}$, leaving group completion aims to identify the corresponding leaving groups, i.e. the sub-node set $\mathcal{V}_{lg}$ of $\mathcal{V}_{hg}$. Thus, we regard the leaving group completion as the search in the leaving group hypergraph. RetroSiG considers choosing one node in the leaving group hypergraph and adding it to the explored node set as an action at each step. In the end, we transform the identified nodes into the SMARTS of leaving groups.

\textbf{State Space:} Similar to search in the product molecular graph, each state $s_t \in S$ is represented as $s_t = \{\mathcal{G}_{p}, \mathcal{G}_{hg}, \hat{\mathcal{V}}_{lg}^{t}\}$, where $\hat{\mathcal{V}}_{lg}^{t}$ denotes the explored node set at step $t$. Notably, $\hat{\mathcal{V}}_{lg}^{t=0} = \emptyset$.

\textbf{Action Space:} Here, RetroSiG also has two types of actions in its action space $\mathcal{A}$: (1) selecting one node $v$ in the leaving group hypergraph $\mathcal{G}_{hg}$, and (2) stop action, (i.e., doing nothing), denoted as $\text{STOP}$. In the evaluation set, 98.9\% of the leaving groups that co-occur within a template are contained within the $\mathcal{E}_{hg}$. It implies that, during inference, the prediction is likely to be incorrect if the chosen leaving group originates from different hyperedges. Thus, we impose a one-hop constraint on the first type of action. This constraint serves to reduce the search space and enhance overall performance. In other words, the action $a_t$ is to select one node $v$ from the first-order neighbour of the current explored node set $\hat{\mathcal{V}}_{lg}^{t}$ or a $\text{STOP}$ action. At step $t$, action space is represented as $\mathcal{A}^{t>0} = \{v | v \in \operatorname{ONE-HOP}(\hat{\mathcal{V}}_{lg}^{t}\} \cup \{\text{STOP} \}$. Notably, $\mathcal{A}^{0} = \{v | v \in \mathcal{V}_{hg} \}$.


\textbf{Reward Function:} At intermediate step $t$, reward function $\mathcal{R}$ gives $s_t$ a reward 0. At the terminal step $T$, if the predicted $\hat{\mathcal{V}}_{lg}^{T}$ is identical to the ground-truth $\mathcal{V}_{lg}$, $\mathcal{R}$ gives $s_t$ a reward 1. Otherwise, it gives $s_t$ a reward of 0.

\textbf{Policy Net (Fig. \ref{policy}b):} We use the HGNN \cite{feng2019hypergraph} to obtain node-level embedding. We assign each atom $u$ in the hypergraph a randomly initialized learnable embedding $\mathbf{f}_u$. To simplify, we represent the encoding process as $\mathrm{HGNN}(\cdot)$ and elaborate on the architectural specifics in Section \ref{preliminaries}. The HGNN computes atom representations $\left\{\mathbf{h}_u^{t} \mid u \in \mathcal{V}_{hg}\right\}$ via
\begin{equation}
\left\{\mathbf{h}_u^{t}\right\}=\operatorname{HGNN}\left(\mathcal{G}_{hg},\left\{\mathbf{f}_u + \mathbb{I}[u \in \hat{\mathcal{V}}_{lg}^{t}] \mathbf{f}_{explored}\right\}\right).
\end{equation}
In order to incorporate information about the explored node set $\hat{\mathcal{V}}_{lg}^{t}$ at step $t$, we add a learnable embedding $\mathbf{f}_{explored}$ to the initial embedding of the explored node. $\mathbb{I}[u \in \hat{\mathcal{V}}_{lg}^{t}] \mapsto\{0,1\}$ be the predicate that indicates whether atom $u$ is the element of $\hat{\mathcal{V}}_{lg}^{t}$. The graph representation $\mathbf{h}_{\mathcal{G}}^t$ is an aggregation of atom representations, i.e. $\mathbf{h}_{\mathcal{G}}^t=\sum_{u \in \mathcal{V}_{hg}} \mathbf{h}_u^t$. We also use the EGAT \cite{kaminski2022rossmann} to obtain graph-level embedding $\mathbf{c}_{\mathcal{G}}$ of the product molecular graph $\mathcal{G}_{p}$. Finally, we calculate the logits $s_u^t \in \mathbb{R}^1, s_s^t \in \mathbb{R}^1$ for selecting one atom $u$ and STOP action at each step $t$ through the fully connected layers:
\begin{equation}
\begin{aligned}
s_u^t & =\boldsymbol{W}_b\left(\sigma\left(\boldsymbol{W}_a\left(\mathbf{h}_u^t+\mathbf{c}_{\mathcal{G}}\right)+\boldsymbol{B}_a\right)\right)-\mathbb{I}\left[u \notin \mathcal{A}^t\right] \infty, \\
s_s^t & =\boldsymbol{W}_b\left(\sigma\left(\boldsymbol{W}_a\left(\mathbf{h}_{\mathcal{G}}^t+\mathbf{c}_{\mathcal{G}}\right)+\boldsymbol{B}_a\right)\right)-\mathbb{I}\left[\mathrm{STOP} \notin \mathcal{A}^t\right] \infty,
\end{aligned}
\end{equation}
where $\mathbb{I}[u \notin \mathcal{A}^{t}] \, (\mathbb{I}[\text{STOP} \notin \mathcal{A}^{t}]) \mapsto\{0,1\}$ be the predicate that indicates whether atom $u$ ($\text{STOP}$) is not in the action space $\mathcal{A}^{t}$ at step $t$. The $-\infty$ operation ensures that the probability of an action not appearing in the action space equals 0 after $\mathrm{SOFTMAX}$ operation.

\subsection{Training and Inference}
The goal of policy network aims to maximize the expected reward, enhance search quality across episodes and it is updated using Proximal Policy Optimization (PPO) \cite{schulman2017proximal}:
\begin{equation}
L^{C L I P}(\theta)=\hat{E}_t[\min (r_t(\theta) \hat{A}_t, \operatorname{clip}\left(r_t(\theta), 1-\epsilon, 1+\epsilon\right) \hat{A}_t)]. 
\end{equation}
Here, $\hat{E}_t$ represents the expected value at timestep t, $r_t(\theta)$ is the ratio of the new policy and the old policy, and $\hat{A}_t$ is the estimated advantage at timestep t.

Inference is executed via beam search \cite{ma2021global} employing a log-likelihood scoring function. For a beam size $k$, we can obtain two sets of $k$ candidate solutions from two stages. Finally, we have $k^2$ candidate solutions. From the $k^2$ possibilities, we select $k$ candidate solutions with the highest cumulative log-likelihood score as the top-k prediction results.

\section{Evaluation}
\subsection{Data}
We utilize the well-established retrosynthesis benchmark dataset, USPTO-50K \cite{schneider2016s}, as the basis for evaluating our approach. This dataset comprises 50,016 reactions with accurate atom-mapping, categorized into ten distinct reaction types. Our methodology aligns with the data partition detailed in \cite{coley2017computer}, wherein we allocate 40,000 reactions for training, 5,000 for validation, and 5,000 for testing. To mitigate potential information leakage inherent in the USPTO-50K dataset, as highlighted in previous studies \cite{somnath2021learning, yan2020retroxpert}, we apply additional steps. This involves canonicalizing the product SMILES and reassigning the atom-mapping to the reactant atoms, following the methodology outlined in \cite{somnath2021learning}.

\subsection{Evaluation Metric}
In line with previous studies, we adopt top-k exact match accuracy as our evaluation metric. In our experiments, we utilized different values of k—1, 3, 5, and 10—for comparative analysis. This metric assesses whether the predicted set of reactants precisely matches the ground truth reactants.

\subsection{Baseline}
We contrast the predictive outcomes of RetroSiG against various template-based, template-free, and semi-template-based methodologies. Notably, we introduce semi-template-based methods including G2G \cite{shi2020graph}, RetroXpert \cite{yan2020retroxpert}, RetroPrime \cite{wang2021retroprime}, MEGAN \cite{sacha2021molecule}, GraphRetro \cite{somnath2021learning}, and Graph2Edits \cite{zhong2023retrosynthesis} as primary baselines due to their exceptional performance. We also considered template-based models like Retrosim \cite{coley2017computer}, Neuralsym \cite{segler2017neural}, GLN \cite{dai2019retrosynthesis}, LocalRetro \cite{chen2021deep}, and template-free models such as SCROP \cite{zheng2019predicting}, Augmented Transformer \cite{tetko2020state}, GTA \cite{seo2021gta}, Graph2SMILES \cite{tu2022permutation}, Dual-TF \cite{sun2021towards}, Retroformer \cite{wan2022retroformer} as robust baseline models for comparison.

\subsection{Implementation Details} \cite{wang2019deep}, DeepHypergraph\footnote{https://github.com/iMoonLab/DeepHypergraph} (DHG), and PyTorch \cite{paszke2019pytorch}. In the EGAT framework, we configure four identical four-head attentive layers with a hidden dimension of 512, while the HGNN structure consists of two layers with a hidden dimension of 512. Embedding sizes across the model are uniformly set at 512. We use $\operatorname{ELU}(x)=\alpha(\exp (x)-1)$ for $x \leq 0$ and $x$ for $x > 0$ as our activation function where $\alpha=1$ as our activation function. Our experiments are executed on a machine equipped with an Intel Xeon 4114 CPU and a Nvidia Titan GPU. Training parameters include a learning rate of 0.0001, 10,000,000 iterations, and the utilization of the Adam optimizer \cite{kingma2014adam}. 
At intervals of 1000 iterations, we store checkpoints to identify the best-performing ones based on evaluation criteria.

\subsection{Performance} 
\textbf{Main Results:} Table \ref{main results} displays the top-k exact match accuracy outcomes obtained from the USPTO-50k benchmark. When the reaction class is unknown, our method achieves a 54.9\% top-1 accuracy which is competitive and only 0.2\% lower than the SOTA (Graph2Edits) result of 55.1\%. However, for k equals 3 and 5, RetroSiG outperforms the performance of Graph2Edits. RetroSiG also beats LocalRetro by a margin of 1.5\% and 0.1\% respectively in top-1 and top-3 accuracy. With the
reaction class given, our method achieves a 66.5\% top-1 accuracy which is only 0.6\% lower than the SOTA (Graph2Edits). For k equals 3, 5 and 10, RetroSiG outperforms the performance of Graph2Edits. We attribute it to the solid prior knowledge carried by the one-hop mechanism.

\begin{table*}
\caption{Top-$k$ accuracy for retrosynthesis prediction on USPTO-50K.}
\label{main results}
\begin{center}
\begin{tabular}{p{4cm}ccccp{2cm}cccc}
\hline
\multirow{3}{*}{\textbf{Model}} & \multicolumn{9}{c}{\textbf{Top-k accuracy (\%)}}                                              \\ \cline{2-10} 
                       & \multicolumn{4}{c}{\textbf{Reaction class unknown}} & & \multicolumn{4}{c}{\textbf{Reaction class known}} \\ \cline{2-10} 
                       & \textbf{k=1}    & \textbf{3}      & \textbf{5}      & \textbf{10}       & 
                       &
                       \textbf{1}      & \textbf{3}      & \textbf{5}      & \textbf{10}       \\ \hline
& \multicolumn{8}{c}{Template-Based Methods}                                                                    \\ \hline
Retrosim               & 37.3   & 54.7   & 63.3   & 74.1   &   & 52.9   & 73.8   & 81.2   & 88.1    \\ \hline
Neuralsym              & 44.4   & 65.3   & 72.4   & 78.9    &  & 55.3   & 76.0   & 81.4   & 85.1    \\ \hline
GLN                    & 52.5   & 69.0   & 75.6   & 83.7     & & 64.2   & 79.1   & 85.2   & 90.0    \\ \hline
LocalRetro             & 53.4   & 77.5   & 85.9   & 92.4    &  & 63.9   & 86.8   & 92.4   & 96.3    \\ \hline
& \multicolumn{8}{c}{Template-Free Methods}                                                                     \\ \hline
SCROP                  & 43.7   & 60.0   & 65.2   & 68.7     &    & 59.0   & 74.8   & 78.1   & 81.1       \\ \hline
Aug.Transformer        & 53.2   & -      & 80.5   & 85.2      &   & -      & -      & -      & -          \\ \hline
GTA                    & 51.1   & 67.6   & 74.8   & 81.6       &  & -      & -      & -      & -          \\ \hline
Graph2SMILES           & 52.9   & 66.5   & 70.0   & 72.9       &  & -      & -      & -      & -          \\ \hline
Dual-TF                & 53.6   & 70.7   & 74.6   & 77.0       &  & 65.7   & 81.9   & 84.7   & 85.9       \\ \hline
& \multicolumn{8}{c}{Semi-Template-Based Methods}                                                               \\ \hline
G2G                    & 48.9   & 67.6   & 72.5   & 75.5        & & 61.0   & 81.3   & 86.0   & 88.7       \\ \hline
RetroXpert             & 50.4   & 61.1   & 62.3   & 63.4  &    & 62.1   & 75.8   & 78.5   & 80.9    \\ \hline
RetroPrime             & 51.4   & 70.8   & 74.0   & 76.1   &      & 64.8   & 81.6   & 85.0   & 86.9       \\ \hline
MEGAN                  & 48.1   & 70.7   & 78.4   & 86.1    &  & 60.7   & 82.0   & 87.5   & 91.6    \\ \hline
GraphRetro             & 53.7   & 68.3   & 72.2   & 75.5     &    & 63.9   & 81.5   & 85.2   & 88.1       \\ \hline
           Graph2Edits            &  \textbf{55.1}      &  77.3      & 83.4       &     \textbf{89.4}         & &   \textbf{67.1}   &      87.5  &    91.5    &    93.8         \\ \hline
                   RetroSiG    &     54.9   & \textbf{77.6}       & \textbf{84.1}       &  89.0  &           &  66.5      &    \textbf{87.9}    &  \textbf{92.0}     &  \textbf{94.1}           \\ \hline
\end{tabular}
\end{center}
\end{table*}

\begin{figure*}
	\centering
	\includegraphics[width=1\linewidth]{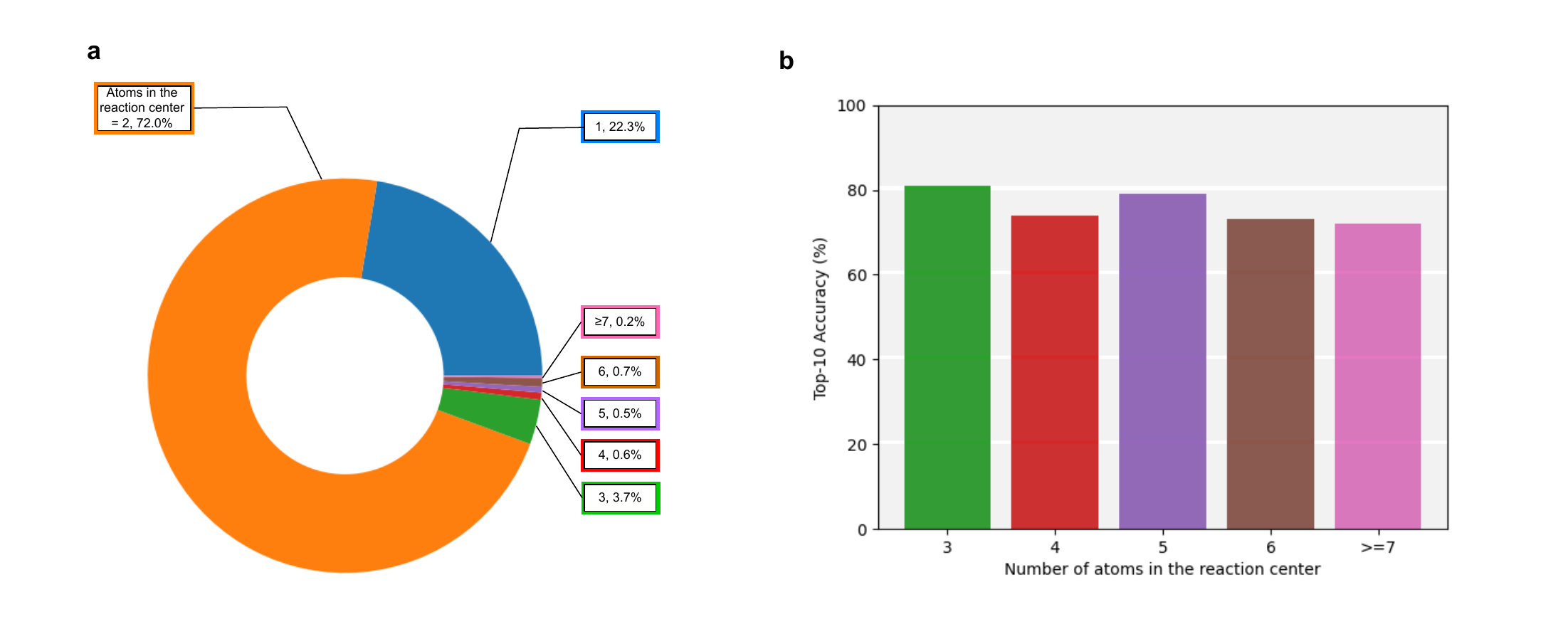}
	\caption{Analysis of complex samples.}
	\label{complex}
\end{figure*}

\textbf{Complex Sample Results:} We investigate the performance effect of some complex reactions in the USPTO-50k, including reactions with multiple reaction center and attaching the same leaving group to more than one atom. As is shown in Fig. \ref{complex}a, Most reactions with 1, 2 atoms in the reaction center account for the majority, which have 1121 (22.3\%) and 3604 (72.0\%) cases respectively. However, the reactions with 3, 4, 5, 6, and $\geq$7 account for a small proportion, which have 184 (3.7\%), 29 (0.6\%), 24 (0.5\%), 36 (0.7\%) and 9 (0.2\%) cases respectively. Multiple reaction center consists of three or more atoms, and 92.6\% (261 / 282) of the samples with multiple reaction center are with attaching the same leaving group to more than one atom. Thus, we report the top-10 accuracy of the number of atoms 3, 4, 5, 6, and $\geq$7 in Fig. \ref{complex}b to investigate the performance of complex reactions. We observe that our model's performance remains stable even as the number of atoms increases. It demonstrates that our model can handle these complex reactions.

\begin{table}[h]
\caption{Ablation Study.}
\label{Ablation Study}
\begin{center}
\begin{tabular}{c|ccc|c}
\hline
\multirow{2}{*}{}  & \multirow{2}{*}{\shortstack{Hypergraph\\Structure}}     & \multirow{2}{*}{\shortstack{One-Hop\\Constraint(rc)}}     & \multirow{2}{*}{\shortstack{One-Hop\\Constraint(lg)}}     &  \multirow{2}{*}{\shortstack{Top-1(\%)}} \\
     & & & & \\
\hline
\textcircled{1} & \CheckmarkBold & \CheckmarkBold & \CheckmarkBold & 54.9  \\
\textcircled{2} & \CheckmarkBold & \CheckmarkBold &   \XSolidBrush                        & 54.6  \\
\textcircled{3} & \CheckmarkBold &    \XSolidBrush                       & \CheckmarkBold & 53.5  \\
\textcircled{4} & \CheckmarkBold &     \XSolidBrush                      &           \XSolidBrush                & 53.4  \\
\textcircled{5} &       \XSolidBrush                    & \XSolidBrush &                    \XSolidBrush       & 52.8  \\ 

\hline
\end{tabular}
\end{center}
\end{table}

\textbf{Ablation Study:} In Fig. \ref{Ablation Study}, Hypergraph-Structure \CheckmarkBold denotes that we use HGNN and hypergraph structure to encode each node (leaving group) embedding. One-Hop Constraint \CheckmarkBold denotes that we use a one-hop constraint. \textcircled{1} outperforms \textcircled{2} and \textcircled{3}, showing that one-hop constraint is effective. \textcircled{2} outperforms \textcircled{3}, demonstrating that one-hop constraint in reaction center identification contributes more to performance improvement. \textcircled{4} outperforms \textcircled{5}, implying that hypergraph effectively models dependencies between leaving groups.

\begin{figure*}
	\centering
	\includegraphics[width=1\linewidth]{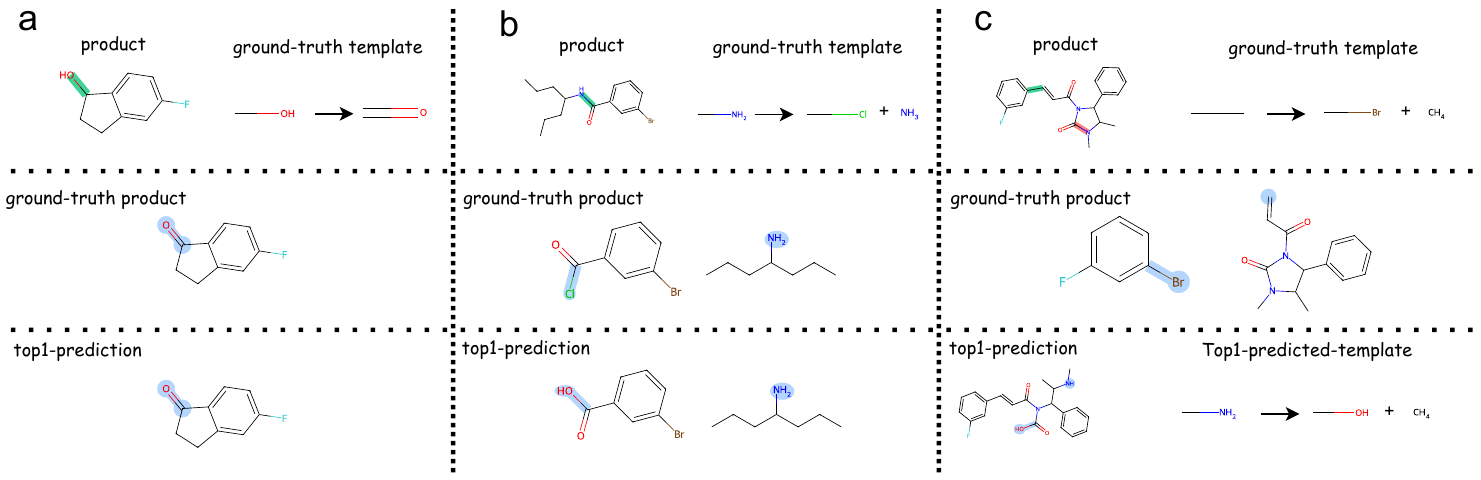}
	\caption{Visualization of Example Predictions.}
	\label{visualization}
\end{figure*}

\subsection{Visualization of Example Predictions}
In Fig. \ref{visualization}, we illustrate the model predictions alongside the corresponding ground truth for three specific cases. Fig. \ref{visualization}a shows an example with attaching the same leaving group to more than one atom. Moreover, the model identifies both the reaction center and leaving groups correctly. Fig. \ref{visualization}b, the model identifies the reaction center correctly, while the predicted leaving groups are incorrect. We hypothesize this is because carboxyl (COOH) is more common in the training set than acyl chloride (COCl). However, the prediction results of top-1 are also feasible chemically. Fig. \ref{visualization}c, the model incorrectly identifies the reaction center and consequently the leaving group, resulting in top-1 prediction being inconsistent with ground truth. Top-1 prediction is an open-loop operation, which is not necessarily harmful from the perspective of multi-step retrosynthesis. Interestingly, the reaction of top-1 prediction is to attach more than one leaving group to one synthon. Previous semi-template-base methods \cite{somnath2021learning} could not predict such complex samples.

\section{Discussion and Conclusion}

\textbf{Limitations:} One common drawback among all semi-template-based and graph-edit methods is their heavy dependence on atom-mapping details between reactants and products. Incorrect mapping will extract the wrong template. It will provide incorrect reaction center labels and leaving group labels during model training. We look forward to the continuous improvement of the atom-mapping algorithm and the extraction/application template algorithm of Rdchiral \cite{coley2019rdchiral} in the community, which is very beneficial to the RetroSiG framework.

\textbf{Future Works:} There are three directions for future work. \textbf{End2End:} We need to design an end-to-end architecture to connect the two search processes so that the signal at the final step can update both stages. \textbf{Decision Transformer :} Our current policy net consists of a simple (hyper) graph encoder and MLP; hence, we must design a more powerful policy net. The ability of Transformer \cite{vaswani2017attention} has recently been proven in various fields, especially in large language models \cite{ouyang2022training}. Naturally, Decision Transformer \cite{chen2021decision} is a choice as a policy net. \textbf{Termination Reward:} Upon completion, the environment currently incentivizes the agent solely when its prediction aligns precisely with the ground truth. However, there exist various valid methods to synthesize a product. It is crucial to devise a more reasonable reward mechanism at termination step.


\bibliographystyle{IEEEtran}
\bibliography{re}

\end{document}